# On the relation between EEG microstates and cross-spectra

Roberto D. Pascual-Marqui[1], Kieko Kochi[1], Toshihiko Kinoshita[2]

1: The KEY Institute for Brain-Mind Research; Department of Psychiatry, Psychotherapy, and Psychosomatics; University of Zurich, Switzerland
2: Department of Neuropsychiatry, Kansai Medical University, Osaka, Japan

**Corresponding author: RD Pascual-Marqui**
**robertod.pascual-marqui@uzh.ch ; https://www.uzh.ch/keyinst**
**scholar.google.com/citations?user=DDqjOkUAAAAJ**

## 1. Abstract

Brain function as measured by multichannel EEG recordings can be described to a high level of accuracy by microstates, characterized as a sequence of time intervals within which the sign invariant normalized scalp electric potential field remains quasi-stable, concatenated by fast transitions. Filtering the EEG has a small effect on the spatial microstate scalp maps, but a large effect on the dynamics (e.g. duration, frequency of occurrence, and transition rates). In addition, spectral power has been found to be strongly correlated with microstate dynamics. And yet, the nature of the relation between spectra and microstates remains poorly understood. Here we show that the multivariate EEG cross-spectrum contains sufficient generative information for estimating the microstate scalp maps and their dynamics, demonstrating an underlying fundamental link between the microstate model and the multivariate cross-spectrum. Empirically, based on EEG recordings from 203 participants in eyes-closed resting state, their cross-spectral matrices were computed, from which stochastic EEG was generated. No significant differences were found for the microstate model (maps and dynamics) estimated from the actual EEG and from the stochastic EEG based solely on the cross-spectra. In addition, with the aim of quantifying the spatio-cross-spectral properties of the microstate model, we introduce here the "topographic likelihood spectrum", based on the Watson distribution, which provides a frequency-by-frequency account of the contribution of a normalized microstate map to the normalized EEG cross-spectrum, independent of power. The topographic likelihood spectra are distinct for the different microstate maps during resting EEG. In a comparison between eyes-closed and eyes open conditions, they are shown to be significantly different in frequency specific patterns.

## 2. Introduction

Brain function as measured by multichannel EEG recordings can be described to a high level of accuracy by microstates, characterized as a sequence of time intervals within which the sign invariant normalized scalp electric potential field remains quasi-stable, concatenated by fast transitions. Up to 70% of explained variance is attained with just four microstate scalp maps, with average duration range 60 to 120 ms. In the early 1970's Lehmann published this phenomenological description based on careful and reproducible visual inspection of multichannel EEG during rest, under an appropriate broad band-pass filter (Lehmann 1971; see also Lehmann et al 1987).

The spatio-temporal aspects of the model are the actual microstate scalp maps, and temporal dynamic parameters (e.g. percent of time in each state, average state duration, number of occurrences per second, and Markov transition rates).

By definition, and due to its very nature, band-pass filtering of the EEG will significantly affect the estimated dynamics, as has been shown empirically (see e.g. Ferat et al 2022). In related research, significant associations have been described between EEG frequency domain parameters and broad-band microstate dynamic parameters (see e.g. Milz et al 2017; Zulliger et al 2022).

In contrast, band-pass filtering has been shown empirically to have little effect on the spatial pattern of the microstate maps (see e.g. Ferat et al 2022).

This present study aims at finding a general relation between the multivariate EEG spectral density and the microstate model. Two main results are presented.

We show that the multivariate EEG cross-spectrum contains sufficient generative information for estimating the microstate scalp maps and their dynamics. This means that original EEG recordings, and stochastic EEGs generated from the cross-spectral matrices, produce microstate scalp maps and dynamics that are statistically similar.

This result is not trivial. This result shows that even the quasi-non-linear aspects of the microstate model can be well approximated as a multivariate linear Gaussian quasi-stationary process, for which the cross-spectrum is known to represent a sufficient statistic. Moreover, this result shows that there is an underlying fundamental link between the microstate model and the frequency domain cross-spectrum.

As a second contribution, we introduce here the "topographic likelihood spectrum" (TLS) as a complimentary descriptor for microstate analyses. The TLS provides a frequency-by-frequency account of the contribution of a normalized microstate map to the normalized EEG cross-spectrum. In other words, it is a measure of alignment of each normalized map with the shape of the space spanned by all the generators of activity at each frequency. A spectral peak in TLS means that at the peak frequency, the map is strongly associated with the predominant normalized generator distribution at that frequency.

Note that the TLS measure is not the power spectrum of an estimator for the microstate signal. Examples of estimators for the microstate signals are of the multivariate type, as in Pascual-Marqui et al (2014) Equation 8 therein, or of the univariate type, as in Ferat et al (2022).

All statistical tests are carried out with non-parametric randomization of the maximum-statistic, thus correcting for multiple testing, and valid even for non-Gaussian data (Karniski et al. 1994; Nichols and Holmes 2002). In addition, and of equal importance, comparisons are also evaluated using effect sizes, following the general guidelines proposed in Poldrack et al (2008).

## 3. Methods

### 3.1. The real EEG data

Babayan et al (2019) have made available in open source form the EEG data used in this study. All details can be found in the original paper, whereas a description of the most important data characteristics can be found in Ferat et al (2022). In summary, the original EEG recording consisted of alternating 60 seconds eyes-closed and 60 seconds eyes open conditions, 16 minutes total, in $N_S = 203$ participants, using $N_E = 61$ scalp electrodes. Preprocessing was performed for artifact correction, with the data downsampled from 2500 Hz to 250 Hz, and band-pass filtered 1 Hz to 45 Hz.

The first 180 seconds of eyes-closed (EC) EEG, and the first 180 seconds of eyes open (EO) EEG, were used here, in the form of $N_W = 180$ epochs of one-second duration each, per condition (EC, EO), per participant. Each one-second epoch consisted of $N_T = 250$ time-samples for $N_E = 61$ electrodes. All analyses in here use the average-reference.

The EEG cross-spectrum of each participant was estimated in the form of the average multivariate periodogram over 180 epochs, using a Hann taper, see e.g. Brillinger (2001), Oppenheim and Schafer (2014), Frei et al (2001). Cross-spectra are computed for all discrete frequencies from 2 Hz to 44 Hz.

The estimated cross-spectrum for the i-th participant is denoted as $\mathbf{S}_i(\omega) \in \mathbb{C}^{N_E \times N_E}$, where $N_E = 61$ is the number of electrodes, and $\omega$ denotes the discrete frequency. Note that the cross-spectral matrices are Hermitian and non-negative definite.

### 3.2. The stochastic EEG data

Stochastic EEG generated from the estimated cross-spectra is obtained as follows. In a first step, random complex Gaussian independent and identically distributed data are generated as $\varepsilon_{ij}(\omega) \sim \mathbf{N}_c(\mathbf{0}, \mathbf{I})$, with $\varepsilon_{ij}(\omega) \in \mathbb{C}^{N_E \times 1}$, for each discrete frequency $\omega$, for the i-the pasrticipant $(i = 1 \ldots N_S)$, for the j-th epoch $(j = 1 \ldots N_W)$. Next, the discrete Fourier transform is computed as:

**Eq. 1** $$\mathbf{Z}_{ij}(\omega) = \left[\mathbf{S}_i^{1/2}(\omega)\right]\left[\varepsilon_{ij}(\omega)\right]$$

where $\mathbf{S}_i^{1/2}(\omega)$ denotes the Hermitian square root of $\mathbf{S}_i(\omega)$. Finally, the inverse Fourier transform of $\mathbf{Z}_{ij}(\omega)$ is computed, giving the real valued stochastic EEG denoted as $\mathbf{Y}_{ij}(t)$, for time sample $(t = 1 \ldots N_T)$. Note that $\mathbf{Z}_{ij}(\omega)$ is $\mathbf{0}$ for frequencies outside the 2-44 HZ range used in this study, except for the conjugate values that are mirrored at frequencies higher than the Nyquist frequency, which guarantees the real-valued $\mathbf{Y}_{ij}(t) \in \mathbb{R}^{N_E \times 1}$.

Eyes-closed stochastic EEGs were generated, matching exactly the sample size of the real eyes-closed EEG recordings used in this study, described above. I.e., 180 seconds of stochastic EC EEG, in the form of $N_W = 180$ epochs of one-second duration each, per participant. Each one-second epoch consisted of $N_T = 250$ time-samples for $N_E = 61$ electrodes.

### 3.3. Microstates: model and estimation

Keeping in mind the phenomenology of ongoing EEG activity that led to the theory of EEG microstates (i.e. brain function as measured by multichannel EEG recordings can be described as a sequence of time intervals within which the sign invariant normalized scalp electric potential field remains quasi-stable, concatenated by fast transitions), the aim here is to use the simplest possible mathematical description of those empirical observations.

For this purpose, the model and corresponding estimation algorithm introduced in Pascual-Marqui et al (1995) is used here, which is aimed at finding the best fitting maps and labels to the EEG measurements. In essence, it is a clustering algorithm of the K-means type, where instead of finding centroids, the method finds the axes in the form of normalized vectors. In summary, the algorithm follows:

Step-A1a: Input scalp maps $\mathbf{X}(t) \in \mathbb{R}^{N_E \times 1}$. If the scalp maps correspond to EEG recordings, then specify the band-pass filter (optional). Alternatively, the maps need not be a time series, e.g. the collection of all the microstate maps for many participants, in which case no band-pass filter takes place.
Step-A1b: Input the number of microstates $N_\mu$.
Step-A1c: Input the maximum number of iterations and value for small epsilon for testing convergence. Convergence occurs if the absolute value of relative change in global explained variance is smaller than epsilon, or if the number of iterations exceeds the defined maximum.
Step-A1d: Must select one of the two options in Step-A1d1 or Step-A1d2:
Step-A1d1: Input user defined initial normalized microstate maps. Could be, e.g., template microstate maps. Go to Step-A2.
Step-A1d2: Program generated initial normalized microstate maps, by randomly selecting maps from input scalp maps followed by normalization. Go to Step-A2.
Step-A2: For each scalp map $\mathbf{X}(t) \in \mathbb{R}^{N_E \times 1}$, compute its microstate label as:

**Eq. 2** $$L(t) = \arg\min_k \left|\mathbf{f}(t,k)\right|^2$$

with:

**Eq. 3** $$\mathbf{f}(t,k) = \mathbf{X}(t) - a\boldsymbol{\Gamma}_k$$

and:

**Eq. 4** $$a = \boldsymbol{\Gamma}_k^T \mathbf{X}(t)$$

where $\left|\mathbf{f}(t,k)\right|^2$ is the error of fit for the model:

**Eq. 5** $$\mathbf{X}(t) = a\boldsymbol{\Gamma}_k + \varepsilon_k$$

Note that $L(t)$ takes integer values in the range 1 to $N_\mu$.
Step-A3: For each microstate label $k = 1 ... N_\mu$, compute the normalized microstate map $\boldsymbol{\Gamma}_k$, as the largest eigenvector of the covariance matrix:

**Eq. 6** $$\mathbf{S}_k = \left[ \frac{1}{\sum_{\forall t : L(t)=k} 1} \right] \left[ \sum_{\forall t : L(t)=k} \mathbf{X}_t \mathbf{X}_t^T \right]$$

Step-A4: Go to Step-A5 if convergence criteria satisfied, otherwise go to Step-A2.
Step-A5: Output: microstate maps $\boldsymbol{\Gamma}_k$ and the labels $L(t)$.

Note that when the input data to the algorithm from Step-A1 to Step-A5 is actual EEG, then the labels $L(t)$ correspond to a time series with possibly "noisy" transitions. In this case, optionally, one can apply Besag's (1986) iterated conditional modes algorithm to the labels $L(t)$. This algorithm will smooth possibly noisy transitions in the label time series, by way of a Markov random field model, which adds a penalized term to the likelihood function, which implements the fact that if $\mathbf{X}(t-1)$ and $\mathbf{X}(t+1)$ both have label "*k*", i.e. $L(t-1) = L(t+1) = k$, then it is highly likely that $\mathbf{X}(t)$ will also have label $L(t)=k$. Note that this method can be extended to larger time windows. A detailed account of the implementation of Besag's algorithm is given in Pascual-Marqui et al (1995).

On occasion, several random re-initializations at Step-A1d2 are performed, and the model parameters corresponding to the highest global explained variance are kept.

As in Ferat et al (2022), five microstates $(N_\mu = 5)$ were used.

A 2-20 Hz band-pass filter was applied within the microstate estimation algorithm to the EEG (for real recordings, and for stochastically generated data).

For each participant and each condition separately, the best fitting microstates were estimated by running the algorithm with 20 random-reinitializations, and 50 maximum iterations.

In a second step, the collection of 406 best fitting normalized microstates (203 participants, in EC and EO conditions), were again submitted to the microstate estimation algorithm. This produced the best overall common best fitting five microstates, reordered to the A,B,C,D,C1 template of Ferat et al (2022).

In a final step, the common microstate maps were used in the algorithm as initialization in Step-A1d1, and all the filtered 2-20Hz EEGs (real recordings and stochastic EEGs) were processed by the algorithm with 100 maximum iterations, giving the final set of microstate parameters, for each participant and condition (EC, EO, real, stochastic). Note that the initialization step with the common microstate maps does not force these maps on the data. It only forces comparable order of estimated maps and parameters, ready for statistical analyses. Without this step, the estimated maps are in arbitrary order.

### *3.4. Microstate parameters*

The microstate parameters computed in this study are described next.

Global explained variance percent:

**Eq. 7**
$$GEV = \frac{\sum_{\forall t} \left[ \mathbf{\Gamma}_{L(t)}^T \mathbf{X}(t) \right]^2}{\sum_{\forall t} \left[ \mathbf{X}^T(t) \mathbf{X}(t) \right]} \bullet 100$$

Explained variance percent by the k-th microstate:

**Eq. 8**
$$EV_k = \frac{\sum_{\forall t : L(t)=k} \left[ \mathbf{\Gamma}_{L(t)}^T \mathbf{X}(t) \right]^2}{\sum_{\forall t} \left[ \mathbf{X}^T(t) \mathbf{X}(t) \right]} \bullet 100$$

Time fraction in the k-th microstate:

**Eq. 9**
$$Time_k = \frac{\sum_{\forall t : L(t)=k} 1}{\sum_{\forall t} 1}$$

The frequency of occurrence per second (i.e. number of occurrences per second) of the k-th microstate is defined as total number of occurrences of the k-th microstate, divided by the total recording time. Referring to Figure 1, there are four occurrences of microstate "A" per "$T$" seconds, three occurrences of microstate "B" per "$T$" seconds, and one occurrence of microstate "C" per "$T$" seconds.

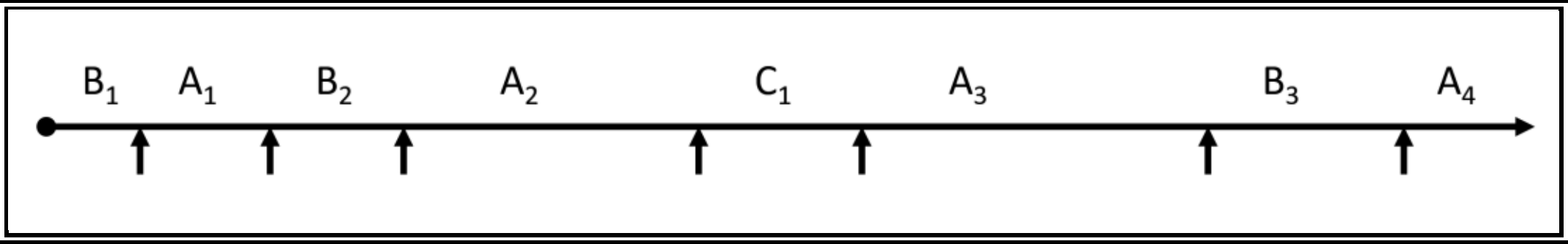

Figure 1: Horizontal axis is time, and total time from start to end duration is "$T$" seconds. Small vertical arrows denote transitions between three microstates denoted A,B,C. The occurrences of each microstate are numbered in the subscript.

The average *duration* of the k-th microstate is another important parameter. For instance, referring to Figure 1, the average duration of microstate "A" is the average of the four durations "$A_1$, $A_2$, $A_3$, and $A_4$". Note that care should be taken at the borders (start and end of each epoch), since the actual durations are always larger. One solution is to simply trim out the start and end microstates and to redefine the start and end of the epoch accordingly.

### *3.5. Rates of transitions between microstates*

In this study, we emphasize that transitions between microstates will not be modeled as a discrete Markov chain, as in vonWenger et al 2017. In our study, it is deemed incorrect to use such a model, because it does not consider the fact that when each state occurs, it does so for a random finite amount of time.

A more appropriate model for describing the transitions is to use a continuous time Markov process model. In this case, the parameters of interest are the transition rates (in number of jumps per unit time) from the i-th to the j-th microstate, with maximum likelihood estimator:

**Eq. 10** $$TransRate(i \to j) = \frac{TotNumbJumps(i \to j)}{TotTimeIn(i)} \quad , \; for\; i \neq j$$

**Eq. 11** $$TransRate(i \to i) = -\sum_{\forall j:\, j \neq i} TransRate(i \to j)$$

where $TotNumbJumps(i \to j)$ denotes the total number of times a transition from $(i \to j)$ occurs, and $TotTimeIn(i)$ denotes the sum total time (in seconds) in the i-th microstate.

See Basawa and Prakasa Rao 1980, Equations 95 to 101 therein.

The self-transition rate $TransRate(i \to i)$ is negative, and from its definition in Eq. 11, it corresponds to the rate of escaping the i-th state. I.e., the more negative, the shorter average duration of the i-th state. While the self-transition rate might seem redundant since it shares very similar information with the "average duration" parameter previously defined, the transition rate matrix in its entirety contains novel information.

We emphasize that even if the transitions do not follow a Markov process, the transition rates constitute an important descriptor for the process. This is in much the same way that the mean and standard deviations are descriptors of data, regardless of the possibility that the data does not have a Gaussian distribution.

### *3.6. The topographic likelihood spectrum (TLS)*

The topographic likelihood spectrum (TLS) for each normalized microstate map $\mathbf{\Gamma}_k$, with $k = 1...N_\mu$, is defined as:

**Eq. 12** $$TLS_k(\omega) = \mathbf{\Gamma}_k^T \mathbf{S}_{tr}(\omega) \mathbf{\Gamma}_k$$

where $\mathbf{S}_{tr}(\omega)$ is the unit trace cross-spectral matrix at frequency $\omega$. This definition is loosely based on the Watson probability density function, see e.g. Mardia and Jupp (1999), Equation 10.3.30 therein.

Note that the global field power spectra for $\mathbf{S}_{tr}(\omega)$ is constant. I.e. it has a flat spectrum, which is devoid of power information. For this reason, the TLS is not related to the power spectra of the microstate maps.

This measure provides a frequency-by-frequency account of the contribution of a normalized microstate map to the normalized EEG cross-spectrum. In other words, it is a measure of alignment of each normalized map with the shape of the space spanned by all the generators of activity at each frequency. A spectral peak in TLS means that at that the peak frequency, the map is strongly associated with the predominant normalized generator distribution at that frequency. In other words, a microstate map has large contribution to a cross-spectral matrix if it is close to some of its largest eigenvectors.

## 4. Results and discussion

### *4.1. Microstate scalp maps*

Figure 2 displays the five best fitting common microstates estimated in this study, based on 3 min EC and 3 min EO EEG for 203 participants, band-pass filtered 2-20 Hz. These microstates are qualitatively similar to those from Ferat et al (2022). This result serves as a sanity check.

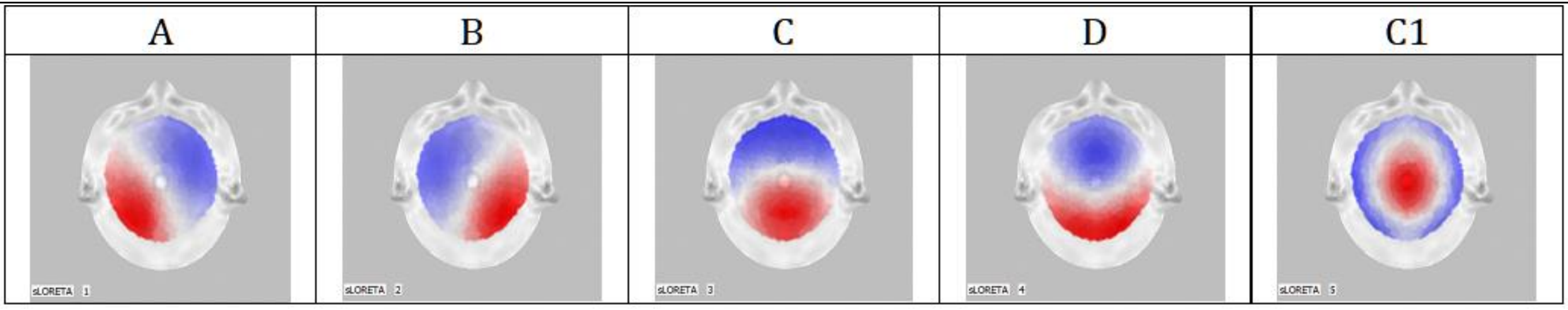

Figure 2: Five best fitting common microstates estimated in this study, based on 3 min EC and 3 min EO EEG for 203 participants, band-pass filtered 2-20 Hz.

### *4.2. Average microstate parameters*

Using the microstate template from Figure 2 as initialization, the best fitting microstates (maps and labels) were computed separately for each participant, and each condition (EC, EO, real recordings, generated stochastically), for band-pass filtered data 2-20 Hz. After convergence allowing for a maximum of 100 iterations, the initialization only guarantees a similar ordering of the individual best fitting microstates. The set of averages of each individual microstate parameter over 203 participants is shown in Table 1.

Table 1: Average microstate parameters over 203 participants.

| Real EEG EC | A | B | C | D | C1 |
|---|---|---|---|---|---|
| %ExpVar | 10.20301 | 11.47279 | 18.24970 | 19.35359 | 6.88059 |
| %TotTime | 0.18024 | 0.19249 | 0.23678 | 0.24425 | 0.14629 |
| #Occurrences | 3.29627 | 3.39412 | 3.76055 | 3.83304 | 2.94099 |
| Duration | 0.05505 | 0.05699 | 0.06287 | 0.06386 | 0.04993 |
| A → (jumps/s) | -17.16952 | 4.15803 | 4.66642 | 4.73112 | 3.61394 |
| B → (jumps/s) | 3.95109 | -16.67175 | 4.58968 | 4.73320 | 3.39778 |
| C → (jumps/s) | 3.64379 | 3.76162 | -15.32156 | 4.84520 | 3.07095 |
| D → (jumps/s) | 3.58529 | 3.78668 | 4.60935 | -15.10457 | 3.12326 |
| C1 → (jumps/s) | 4.42333 | 4.58070 | 4.91178 | 4.87723 | -18.79304 |
| | | | | | |
| **Real EEG EO** | **A** | **B** | **C** | **D** | **C1** |
| %ExpVar | 10.69300 | 11.24069 | 16.77654 | 16.26818 | 8.08635 |
| %TotTime | 0.19201 | 0.19657 | 0.22465 | 0.22038 | 0.16633 |
| #Occurrences | 3.72493 | 3.77419 | 4.02176 | 4.03406 | 3.43580 |
| Duration | 0.05170 | 0.05228 | 0.05564 | 0.05457 | 0.04829 |
| A → (jumps/s) | -18.36254 | 4.67403 | 4.84876 | 4.86977 | 3.96998 |
| B → (jumps/s) | 4.54369 | -18.15548 | 4.82022 | 4.87851 | 3.91307 |
| C → (jumps/s) | 4.17862 | 4.29516 | -17.16527 | 4.86965 | 3.82185 |
| D → (jumps/s) | 4.29133 | 4.34176 | 4.85935 | -17.41136 | 3.91893 |
| C1 → (jumps/s) | 4.60452 | 4.63003 | 5.22210 | 5.08849 | -19.54514 |
| | | | | | |
| **Stochastic EEG EC** | **A** | **B** | **C** | **D** | **C1** |
| %ExpVar | 10.69300 | 11.24069 | 16.77654 | 16.26818 | 8.08635 |
| %TotTime | 0.19201 | 0.19657 | 0.22465 | 0.22038 | 0.16633 |
| #Occurrences | 3.72493 | 3.77419 | 4.02176 | 4.03406 | 3.43580 |
| Duration | 0.05170 | 0.05228 | 0.05564 | 0.05457 | 0.04829 |
| A → (jumps/s) | -18.36254 | 4.67403 | 4.84876 | 4.86977 | 3.96998 |
| B → (jumps/s) | 4.54369 | -18.15548 | 4.82022 | 4.87851 | 3.91307 |
| C → (jumps/s) | 4.17862 | 4.29516 | -17.16527 | 4.86965 | 3.82185 |
| D → (jumps/s) | 4.29133 | 4.34176 | 4.85935 | -17.41136 | 3.91893 |
| C1 → (jumps/s) | 4.60452 | 4.63003 | 5.22210 | 5.08849 | -19.54514 |

EC: eyes-closed; EO: eyes open;
%ExpVar: Percent (range 0…100) of explained variance per microstate;
%TotTime: fraction (range 0…1) of time spent in each microstate;
#Occurrences: number of occurrences per second;
Duration: average duration in seconds;
A,B,C,D,C1→: transition rates (row microstate to column microstate) in jumps per second.

### *4.3. Comparisons between microstate parameters for real EEG recordings and EEG generated stochastically from cross-spectra*

Comparisons between EC microstate parameters for real EEG recordings and EEG generated stochastically from cross-spectra are shown in Table 2. For the sample size of 203 participants, the effect size values for Cohen's d, and their equivalent univariate t-statistic values are (small, d=0.2, t=2.84), (medium, d=0.5, t=7.09), (large, d=0.8, t=11.34). No significant differences were found, and all effect sizes were below medium.

Table 2: Univariate t-statistics corresponding to microstate parameters for EC data, comparison for "EEG generated stochastically from cross-spectra" minus "Real EEG". Two-sided threshold corrected for multiple testing t(p<0.01)=3.68 and t(p<0.05)=3.18. Effect sizes, in t-statistic units, are (small, t=2.84), (medium, t=7.09), (large, t=11.34).

| | A | B | C | D | C1 |
|---|---|---|---|---|---|
| %ExpVar | 1.45529 | -0.50829 | 0.88864 | 1.71394 | -1.70096 |
| %TotTime | 2.08715 | -0.10456 | -0.57180 | -0.27868 | -1.27310 |
| #Occurrences | 2.96581 | 2.04987 | 1.86102 | 2.36739 | -1.16678 |
| Duration | -0.18938 | -2.38827 | -1.84294 | -1.73776 | -1.67356 |
| A → (jumps/s) | -0.66558 | -1.36185 | 1.52756 | 2.11191 | -0.94661 |
| B → (jumps/s) | 0.09721 | -2.40456 | 2.37303 | 2.11064 | -0.19734 |
| C → (jumps/s) | 1.69093 | 2.29249 | -1.69932 | 0.48441 | -1.93839 |
| D → (jumps/s) | 2.38064 | 1.35306 | -0.13874 | -1.49289 | -1.00318 |
| C1 → (jumps/s) | 2.80010 | 1.20119 | -0.11888 | -0.46189 | -2.16266 |

t-statistics for:
%ExpVar: Percent (range 0…100) of explained variance per microstate;
%TotTime: fraction (range 0…1) of time spent in each microstate;
#Occurrences: number of occurrences per second;
Duration: average duration in seconds;
A,B,C,D,C1→: transition rates (row microstate to column microstate) in jumps per second.

The meaning of the results in Table 2 can be illustrated as in Figure 3.

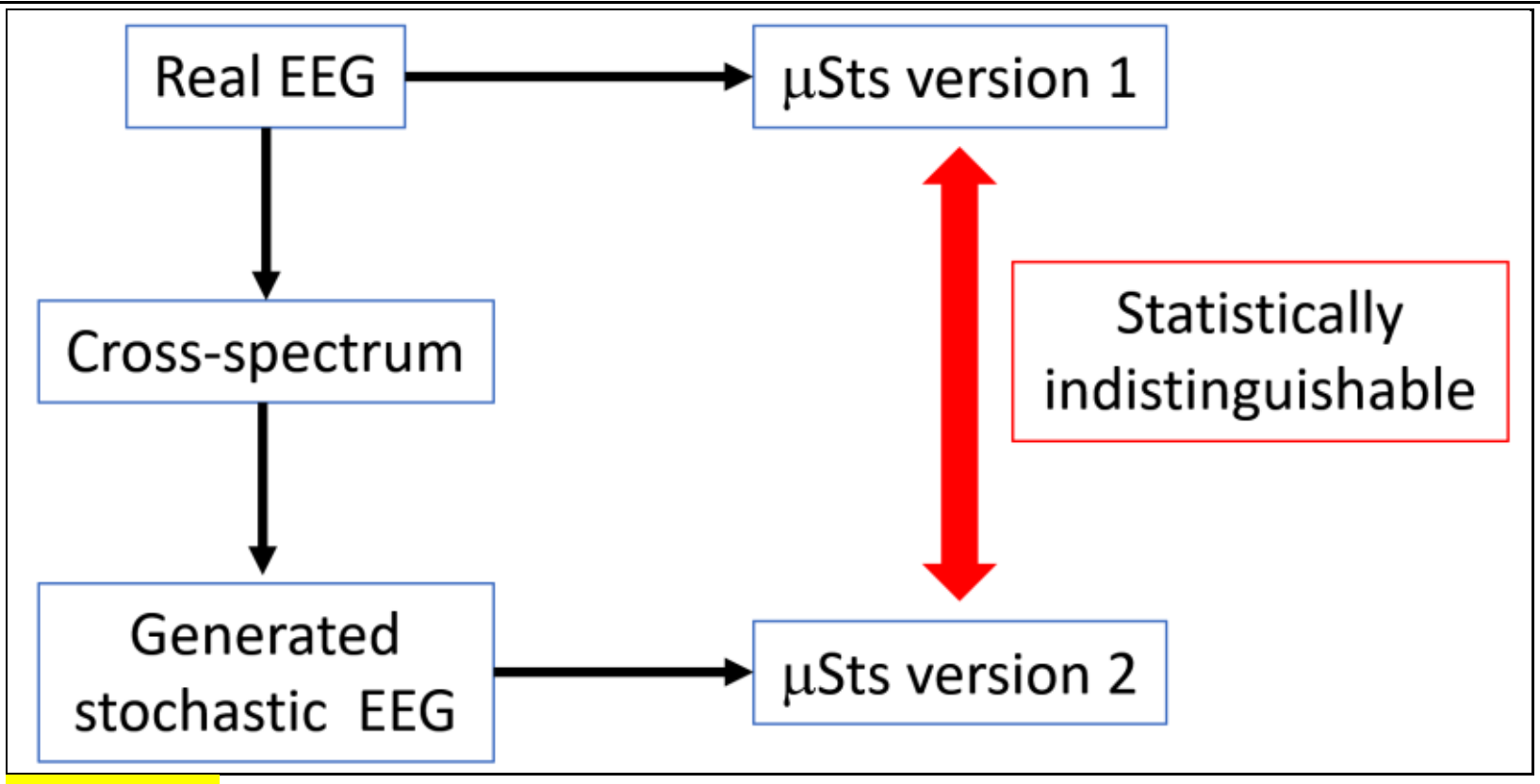

Figure 3: Illustration showing that microstate analysis on real EEG recordings, and on stochastically generated EEG from the cross-spectrum, produce results that are statistically indistinguishable.

### *4.4. Comparisons between microstate parameters for eyes-closed and eyes open conditions*

Table 3 reports the statistical comparisons for microstate parameters for “EC” minus “EO”, for the real EEG recordings. As expected, there are many significant differences between the two states (EC vs EO).

Table 3: Univariate t-statistics corresponding to microstate parameters for “EC” minus “EO”, for the real EEG recordings. Two-sided threshold corrected for multiple testing t(p<0.01)=3.741 and t(p<0.05)=3.202. Effect sizes, in t-statistic units, are (small, t=2.84), (medium, t=7.09), (large, t=11.34). The threshold used for shading corresponds to medium effect size for abs(t)>7.09, which is very much higher than the p<0.01 threshold. Red colors for EC>EO, and green colors for EO>EC.

| | A | B | C | D | C1 |
|---|---|---|---|---|---|
| %ExpVar | -2.16101 | 0.90994 | 3.90048 | 7.57382 | -7.41242 |
| %TotTime | -4.59742 | -1.56888 | 4.19594 | 7.69608 | -9.06096 |
| #Occurrences | -11.57645 | -11.11221 | -8.45636 | -7.12278 | -13.24366 |
| Duration | 8.64417 | 10.27788 | 11.15956 | 13.38529 | 4.76331 |
| A → (jumps/s) | 9.92425 | -7.79988 | -2.80870 | -2.28500 | -6.21937 |
| B → (jumps/s) | -8.88815 | 11.50601 | -3.47705 | -2.45439 | -8.82084 |
| C → (jumps/s) | -8.28920 | -7.98449 | 12.26892 | -0.35757 | -11.73751 |
| D → (jumps/s) | -10.75634 | -9.14987 | -3.89063 | 14.91131 | -12.36941 |
| C1 → (jumps/s | -2.84245 | -0.73063 | -4.51302 | -3.02357 | 5.89356 |

t-statistics for:
%ExpVar: Percent (range 0…100) of explained variance per microstate;
%TotTime: fraction (range 0…1) of time spent in each microstate;
#Occurrences: number of occurrences per second;
Duration: average duration in seconds;
A,B,C,D,C1→: transition rates (row microstate to column microstate) in jumps per second.

As a sanity check, note that the t-values in Table 3, comparing EC minus EO, for %ExpVar (percent of explained variance per microstate), %TotTime (fraction of time spent in each microstate), and for average duration in seconds, are in the same direction as those reported in Ferat et al 2022. These are the parameters which are common to both studies.

Furthermore, it is important to emphasize that the qualitative agreement between these results and those of Ferat et al (2022) occurs despite many methodological differences, such as difference in broad-band limits, and the use of EEG maps at global field power peaks in Ferat et al (2022).

With respect to the continuous time Markov process parameters, eyes open condition has significantly higher transition rates than eyes-closed condition, which means that transitions are more frequent during eyes open leading more often to microstates A, B, and C1.

### *4.5. Topographic likelihood spectra (TLS)*

The average TLS, over 203 participants, for eyes-closed and eyes open conditions are shown in Figure 4 and Figure 5, respectively. Note that TLS was calculated from the cross-spectrum of each of 203 participants for both eyes open and closed conditions, using the common five microstates (Figure 2).

From Figure 4 for EC condition, microstates C and D have the highest spatio-spectral contribution for all frequencies range 2-44 Hz. This is followed by lower spatio-spectral contribution from microstates A and B. And finally microstate C1 has the lowest spatio-spectral contribution. The two strongest spatio-spectral contributors, C and D, have dissociated alpha range peaks, with D for low alpha (around 9 Hz), and C for high alpha (around 11 Hz). In addition, C and D display local maxima close to 20 Hz. A and B have similar spectra, but with peaks (A at 10 Hz) and (B at 12.5 Hz). Microstate C1 peaks at around 3 Hz.

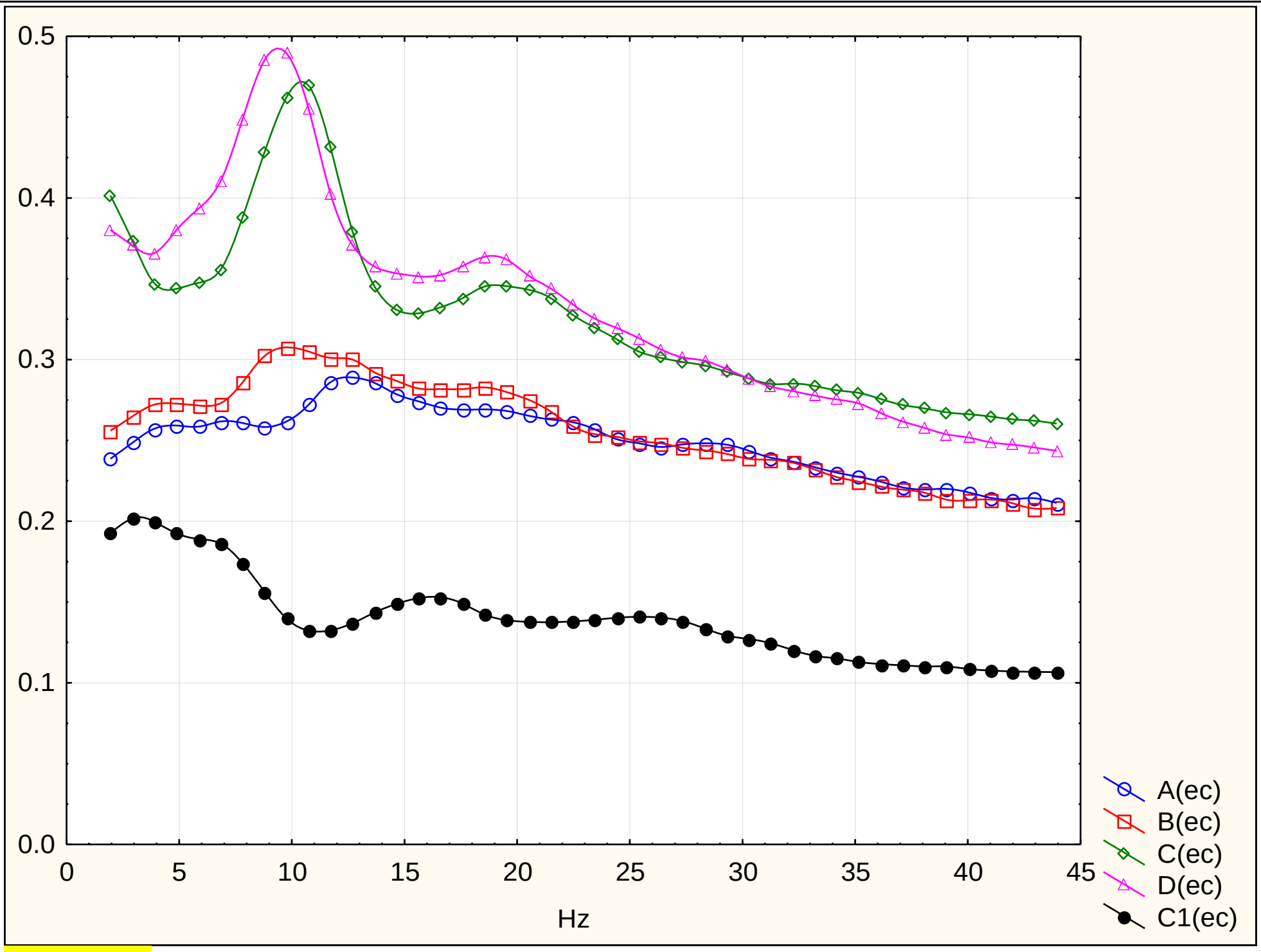

Figure 4: Average topographic likelihood spectra (TLS), over 203 participants, for eyes-closed. TLS is computed from Eq. 12, with the five microstates (Figure 2) common to all 203 participants for both eyes open and closed data.

From Figure 5 for EO condition, microstates C and D have the highest spatio-spectral contribution for all frequencies range 2-44 Hz. This is followed by lower spatio-spectral contribution from microstates A and B. And finally microstate C1 has the lowest spatio-spectral contribution. The two strongest spatio-spectral contributors, C and D, have dissociated peaks, with D displaying a 7 Hz theta peak, and C with a 10 Hz alpha peak. In addition, D displays a local maximum close to 14 Hz. A and B have similar spectra, but with peaks at 11 to 12 Hz. Microstate C1 peaks at around 3 Hz.

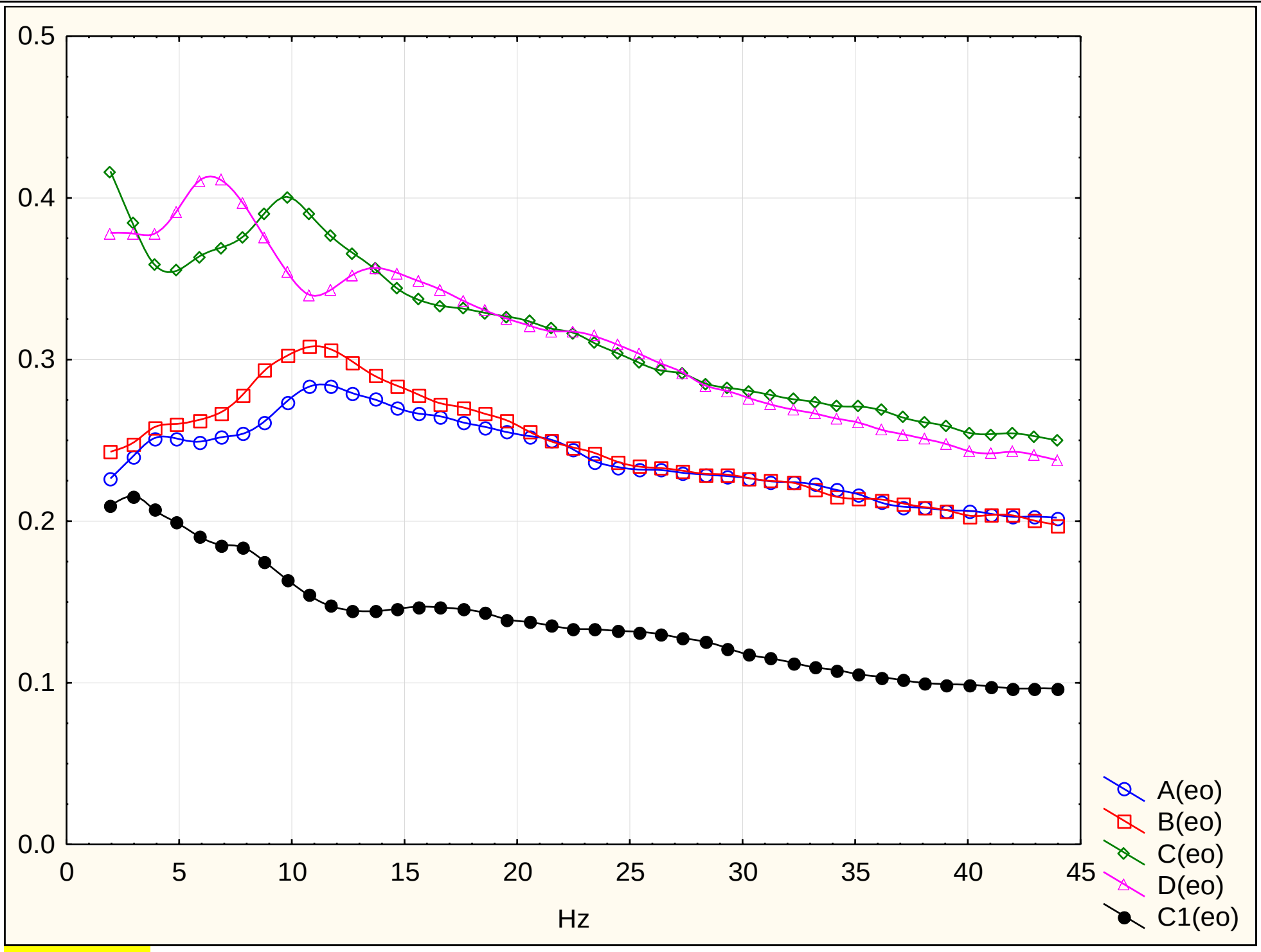

Figure 5: Average topographic likelihood spectra (TLS), over 203 participants, for eyes-open. TLS is computed from Eq. 12, with the five microstates (Figure 2) common to all 203 participants for both eyes open and closed data.

The statistical comparison of TLS for EC minus EO is shown in Figure 6. The vertical axis corresponds to t-value units, and the horizontal axis to Hz. Two-sided thresholds corrected for multiple testing are t(p<0.01)=4.056 and t(p<0.05)=3.595. Effect sizes, in t-statistic units, are (small, t=2.84), (medium, t=7.09), (large, t=11.34).

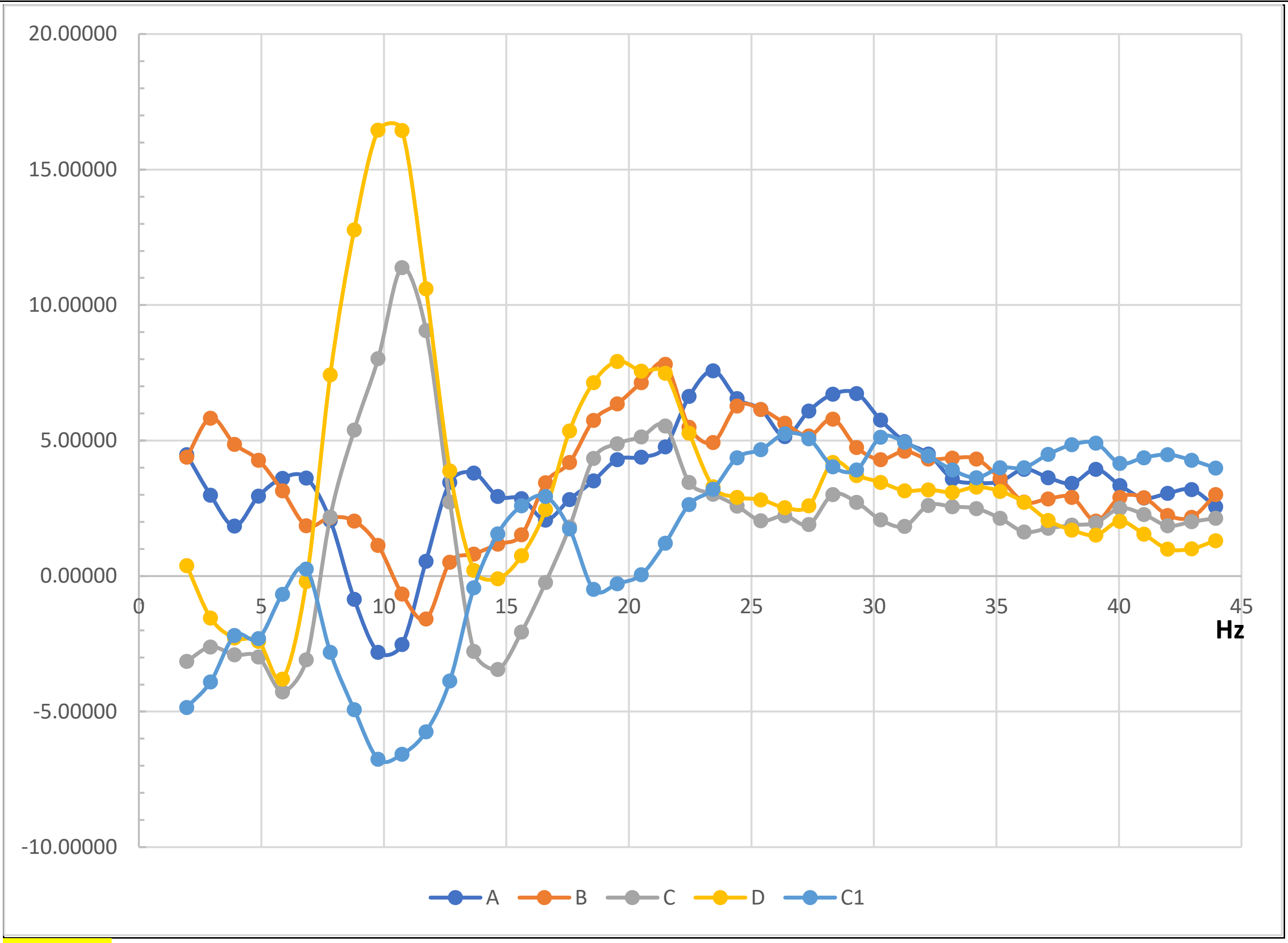

Figure 6: Statistical comparison of TLS for EC minus EO. The vertical axis corresponds to t-value units, and the horizontal axis to Hz. Two-sided thresholds corrected for multiple testing are t(p<0.01)=4.056 and t(p<0.05)=3.595. Effect sizes, in t-statistic units, are (small, t=2.84), (medium, t=7.09), (large, t=11.34).

The major significant features of the comparison for TLS(EC-EO) are:
1. Microstate C1 in eyes open condition has larger topographic likelihood at 10 Hz.
2. Microstates C and D in eyes-closed condition have larger topographic likelihood at 10 Hz.
3. In eyes-closed condition, microstates A, B, C, and D have larger topographic likelihood at around 20 Hz.

## 5. Concluding summary

It was shown here empirically that microstate model parameters (maps and dynamics) can be estimated from stochastic EEG generated from its multivariate cross-spectrum. It is important to note that:
1. This result is not trivial. This result shows that even the quasi-non-linear aspects of the microstate model can be well approximated as a multivariate linear Gaussian quasi-stationary process, for which the cross-spectrum is known to represent a sufficient statistic. Moreover, this result shows that there is an underlying fundamental link between the microstate model and the frequency domain cross-spectrum.
2. The cross-spectrum is a sufficient statistic for linear, quasi-stationary, Gaussian signals. Even if linearity and Gaussianity are violated, the cross-spectrum is still very informative, since it corresponds to the multivariate second order moment. This seems to suffice for reproducing the microstate model.
3. Even though cross-spectra are embedded with microstate information, they are no substitute for the microstate model. Cross-spectra are difficult to interpret in their raw form, while the microstate model provides a set of meaningful, interpretable features.

The topographic likelihood spectrum (TLS) constitutes a quantification of the spatio-cross-spectral properties of the microstate model. It provides a frequency-by-frequency account of the contribution of a normalized microstate map to the normalized EEG cross-spectrum, independent of power. In other words, it is a measure of alignment of each normalized map with the shape of the space spanned by all the generators of activity at each frequency. A spectral peak in TLS means that at the peak frequency, the map is strongly associated with the predominant normalized generator distribution at that frequency.

Different microstate maps have different TLS, which are also different in different brain states. This motivates its use as a possible biomarker.

Again: the TLS measure is not the power spectrum of the microstate signal.

## 6. Acknowledgements

We are indebted to Dr Tomas Ros for his time and effort in converting the open access EEG recordings used here into human readable text format, thus making it accessible to analysis with the LORETA-KEY free academic software (https://www.uzh.ch/keyinst/).